\documentclass[conference]{IEEEtran}
\usepackage[dvipdfmx]{graphicx}
\usepackage[dvipdfmx]{color}
\usepackage{bm}
\usepackage{amssymb}
\usepackage{amsmath}
\usepackage{newtxtext}
\usepackage{caption}
\newcommand{\sato}{}

\newcommand{\shimo}{}
\newcommand{\sntn}{}

\newcommand{\mr}[1]{\mathrm{#1}}
\newcommand{\mb}[1]{\mathbf{#1}}
\newcommand{\mca}[1]{\mathcal{#1}}
\newcommand{\bs}[1]{\boldsymbol{#1}}
\newcommand{\X}{\mb{X}}
\newcommand{\Y}{\mb{Y}}

\newcommand{\K}{\mb{K}}

\newcommand{\f}{\mb{f}}
\newcommand{\W}{\mb{W}}

\title{
Adaptive Outlier Detection for Power MOSFETs
Based on Gaussian Process Regression
}

\author{
    Kyohei Shimozato$^\mathsection$
    Michihiro Shintani$^\dagger$, and
    Takashi Sato$^\mathsection$\\\\
    $^\mathsection$ Graduate School of Informatics, Kyoto University \quad
    Yoshida-hon-machi, Sakyo, Kyoto 606-8501, Japan \\
    $^\dagger$ Graduate School of Science and Technology, Nara Institute of Science and
    Technology \\
    8916-5 Takayama-cho, Ikoma, Nara 630-0192, Japan \\
    Phone: +81-75-753-4801 \quad E-mail: paper@easter.kuee.kyoto-u.ac.jp \\
}

\begin{document}

\maketitle


\begin{abstract}
Outlier detection of semiconductor devices is important since
manufacturing variation is inherently inevitable.  In order to properly
detect outliers, it is necessary to consider the discrepancy from
underlying trend. Conventional methods
are insufficient {\sato{as they cannot track \shimo{spatial} changes of the
trend}}. This study proposes an adaptive outlier detection using
Gaussian process regression (GPR) with Student-t likelihood,
{\sato{which captures}} a gradual spatial change of characteristic
variation. According to the credible interval of the GPR posterior
distribution, the devices having excessively large deviations against
the underlying trend are detected.  The proposed methodology is
validated by the experiments using a commercial SiC wafer {\sato{and
simulation}}.
\end{abstract}
\vspace{\baselineskip}
\begin{IEEEkeywords}
  Outlier detection, characteristic variation, Gaussian process regression
\end{IEEEkeywords}

\section{INTRODUCTION}
Power devices, such as SiC MOSFETs,
are the important components for building efficient converters.
Semiconductor devices are subject to characteristic variations, and
power devices
are no exception.
Typically, the characteristic variation of a chip on a wafer
is known to be separated into {\sato{two:}} an underlying trend and a random noise added to it~\cite{stine_analysis_1997}.
Therefore, the proper modeling of fundamental trends and the decomposition of these components 
are important for {\sato{detecting outliers and improving manufacturing process.}}

SiC wafers are commonly manufactured {\sato{using}} physical vapor transport (PVT) {\sato{crystal growth}}.
In the PVT method, wafers {\sato{are grown}} in a heated crucible,
so the basal plane bending and crystallographic dislocations
{\sato{due}} to temperature gradients are unavoidable.
This heterogeneity is considered as the cause of
{\sato{the spatial variation of the characteristics}}~\cite{kimoto_bulk_2016,cui_spatial_2017,mahadik_evolution_2020}.
In addition, bulk micro-defects
can be formed randomly.
{\sato{Numerous studies have}} reported that {\sato{such}} random defects can pose a
reliability risk~\cite{hasegawa_which_2020,madge_statistical_2002},
{\sato{thus chips that may contain}} these defects must be judged as outliers.
Chips with excessively large deviation from the {\sato{spatial trend are likely to contain such defects
because these random defects have a significant impact on the characteristic degradation.}}

{\sato{In practice, it is difficult to detect}} these anomalies
{\sato{using}}
conventional {\sato{methods.  Among others, the dynamic part average
testing (DPAT)~\cite{PAT} and nearest neighbor residual
(NNR)~\cite{daasch_variance_2000,sabade_comparison_2004} are widely used
testing methods.}}  DPAT is based on wafer-wide distribution and hence
cannot capture the chips whose characteristics deviate significantly
from the
\shimo{spatial} trend.  NNR, on the other hand, takes into
account {\sato{the local trend changes.  It predicts the trend
based on the characteristics of neighboring}} chips.  However, since the
prediction is carried out using the limited number of neighboring chips,
the trend may be {\sato{biased by the presence of a cluster of outliers.}}

In this study, adaptive detection of outlier chips
based on a statistical methodology is proposed.
The Gaussian process regression (GPR)~\cite{rasmussen_gaussian_2005,PRML_2006}
with Student-t likelihood~\cite{vanhatalo_gaussian_2009} is utilized
to define {\sato{both}} the {\sato{spatial characteristic trend}} and the allowable range {\sato{of the characteristic}}.
GPR is a non-parametric statistical model {\sato{that can 
calculate the underlying trends from}} the measured data {\sato{without}} prior assumption {\sato{of the model function}}.
{\sato{Through GPR}}, the credible interval can be obtained in addition to the mean of the prediction.
Since this interval is {\sato{an indicator for finding}} a statistically reasonable range of the chip characteristics,
{\sato{we use the interval to judge the outliers.}} 

The {\sato{proposed outlier detection method}} is validated
{\sato{using}} the wafer measurements of a commercial power MOSFET and
an artificially generated dataset {\sato{that we know the ground
truth.  The performance of the detection}} is
compared with that of conventional methods.
{\sato{The experimental results show that the accuracy of the proposed
method is significantly better than}} that of the conventional methods.


\begin{figure*}[]
  \centering
  \includegraphics[width=0.8\linewidth]{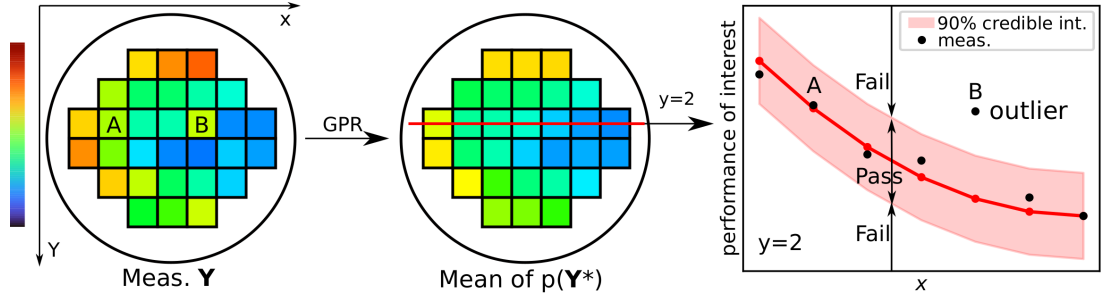}
  \caption{Concept of the proposed adaptive outlier detection.}
  \label{fig:flow}
\end{figure*}

\section{GAUSSIAN PROCESS REGRESSION}

In this section, the details of Gaussian process regression (GPR) are explained.
GPR is a regression method based on the Gaussian process {\sato{and}} 
can be adapted {\sato{without assuming the form of the regression function beforehand.}}
From the known input-output pairs $\X \to \Y$ of a latent function $f$,
the output $Y^*$ {\sato{that}} corresponds to arbitrary inputs $X^*$ can be predicted using GPR.
In GPR, the prediction $Y^*$ is obtained as a probability distribution,
which is useful {\sato{for determining}} whether the regression results are {\sato{credible}} or not.
{\sato{The}} narrower the distribution as a prediction, {\sato{the greater the certainty,
whereas the wider the distribution, the greater the likelihood that the actual observation may contain larger randomness.}}

Assume that $\f = (f(X_1), f(X_2),..., f(X_n))$ as a set of random variables
for inputs $\X = (X_1, X_2,..., X_n)$.
When $\f$ {\sato{is}} a Gaussian process (GP), any finite subset follows
a multivariate Gaussian distribution
$p(\f) = \mca{N}(\bs{\mu}, \K)$.
Generally, the mean $\bs{\mu}$ is set to $\mb{0}$ and
the covariance matrix $\K = (k_{ij})$ is calculated by a kernel function of $\X$.
A typical kernel function is automatic relevance detection
radial basis function (ARD-RBF):
\begin{align}
  k_{ij} = k(X_i, X_j) = \theta_1 \exp\left(-\sum_{p}\frac{|X_i^{(p)} - X_j^{(p)}|^2}{\theta_2^{(p)}}\right)
  \label{eq:rbf}
\end{align}
where $p$ indicates each dimension of $X$.
$\theta_1, \theta_2$ are hyperparameters.
A set of observation $\Y$ contains random noise in addition to $\f$.
In case that the random noise follows the Student-t distribution,
\begin{align}
  p(Y_i|f_i) &= \mca{ST}(f_i, \nu, \sigma)\nonumber\\
  &=\cfrac{\Gamma((\nu + 1)/2)}{\Gamma(\nu/2)\sqrt{\nu\pi}\sigma}
    \left(1+\cfrac{(Y_i - f_i)^2}{\nu\sigma^2}\right)^{-(\nu + 1)/2}
\end{align}
where $\nu$ is called the degrees of freedom and $\sigma$ the scale parameter.
Since the Student-t distribution has heavier tails compared to
the Gaussian distribution, it is {\sato{tolerant to outliers. Specifically,}}
GPR with Student-t noise is more robust against the outliers
compared to that with Gaussian noise~\cite{vanhatalo_gaussian_2009}.
{\sato{This property is important to ensure that the regression results are
not overly influenced by outliers.}} 

In order to deal with Student-t distribution, Laplace approximation is introduced.
\begin{align}
  p(\f|\Y) &\approx \mca{N} (\f|\hat{\f}, \bs{\Sigma}) \\
  \hat{\f} &= {\mathop{\rm argmax}\limits}_f p(\f|\Y) \\
  \bs{\Sigma} &= \K^{-1} + \W \\
  \W_{ij} &=
    \begin{cases}
      -(\nu + 1)\cfrac{(Y_i-f_i)^2 - \nu \sigma^2}{((Y_i-f_i)^2 + \nu \sigma^2)^2} & (i = j) \\
      0 & (i\ne j)
    \end{cases}
\end{align}
All hyperparameters $\bs{\theta} = (\theta_1, \theta_2, \nu, \sigma)$
are optimized by maximizing log likelihood function
$\mca{L}$~\cite{vanhatalo_gaussian_2009}.
\begin{align}
  \mca{L} &= \log p(\Y|\X,\boldsymbol{\theta})\nonumber\\
  &= \log p(\Y|\hat{\f})
  - \frac{1}{2}\log|{\K}| - \frac{1}{2}\hat{\f}^T\K^{-1}\hat{\f}\\
  &\quad- \frac{1}{2}\log|\K^{-1}+\W| \label{eq:l1}
\end{align}
In practice, gradient-based methods, such as L-BFGS-B method, can be {\sato{applied}} 
for the optimization.

The posterior distribution $p(f^*|X^*)$ for an arbitrary input $X^*$
can be predicted as Gaussian distribution,
because the concatenated vector $(\hat{\f}, f^*)$ also follows
Gaussian process~\cite{vanhatalo_gaussian_2009}.
\begin{align}
  p\left(\begin{pmatrix}
    \hat{\f} \\
    f^*
  \end{pmatrix}\right) &= \mca{N} \left(\mb{0},\,\begin{pmatrix}
    \K & \K_* \\
    \K_*^T & k_{**}
  \end{pmatrix}
  \right) \label{eq:pred1}\\
  p(f^*|\X^*,\X,\hat{\f}) &= \mca{N} (\K_*^T\K^{-1}\hat{\f}, \mb{k_{**} - \mb{k}_*^T\K^{-1}\mb{k}_*}) \label{eq:pred2}
\end{align}
where $\K_{*,i} = k(X_i, X^*)$, $k_{**} = k(X^*, X^*)$.
Then, the prediction of the observation $p(Y^*)$ can be obtained by integrating Student-t distribution,
which can be calculated by quadrature integration.
\shimo{
\begin{align}
  p(Y^*) = \int p(Y^*|f^*)p(f^*|\X^*,\X,\hat{\f})df^*
\end{align}
}
\shimo{For implementing the proposed method,
GPy~\cite{gpy2014}, a Python library of Gaussian process, is {\sato{used}}.}


\begin{figure*}[!t]
  \centering
  \begin{minipage}{0.28\linewidth}
    \centering
    \includegraphics[width=\linewidth]{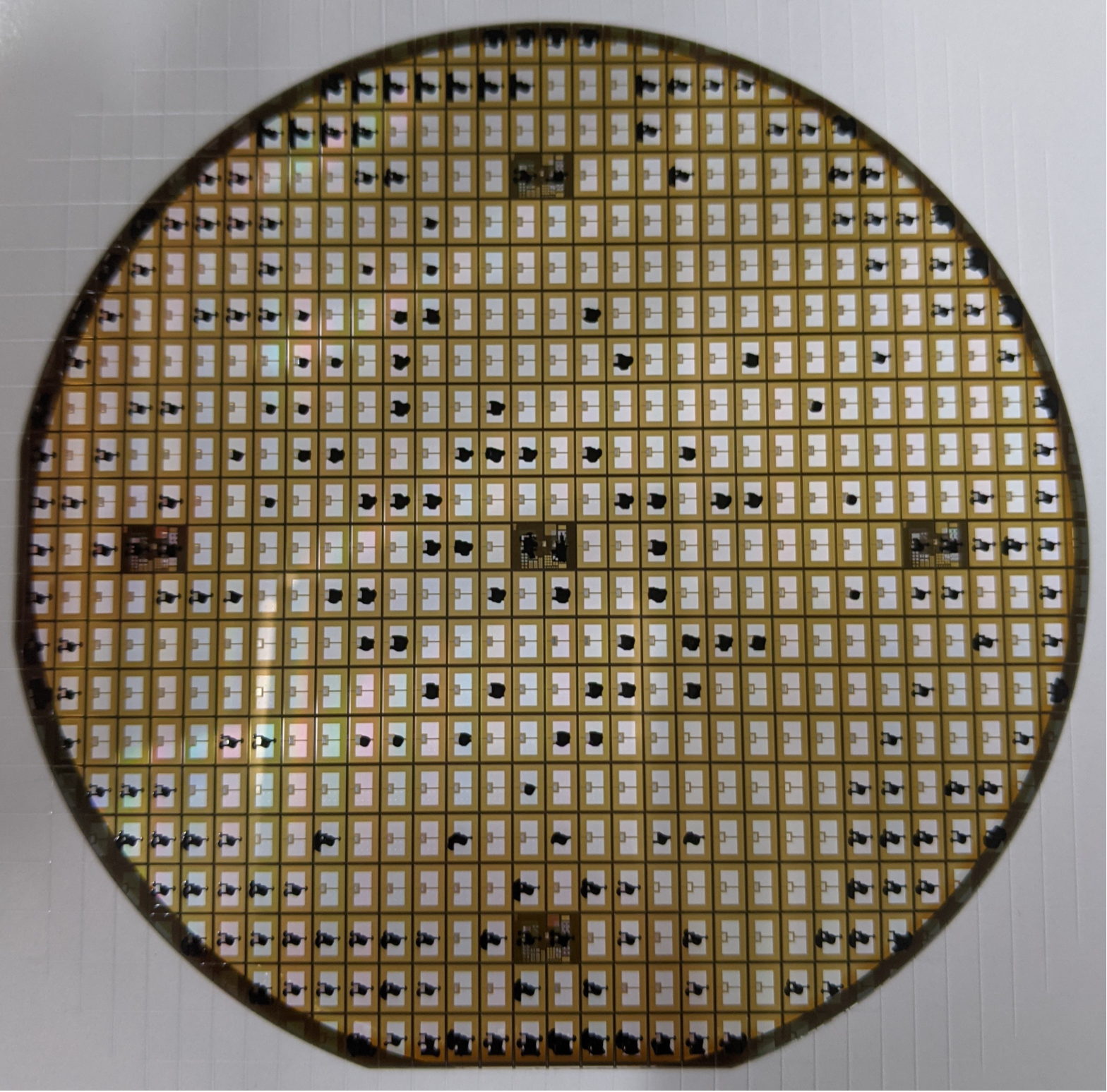}
    \caption{Measured wafer of a commercial SiC MOSFET.}
    \label{fig:s2301}
    \vspace{4truemm}
    \centering
    \includegraphics[width=\linewidth]{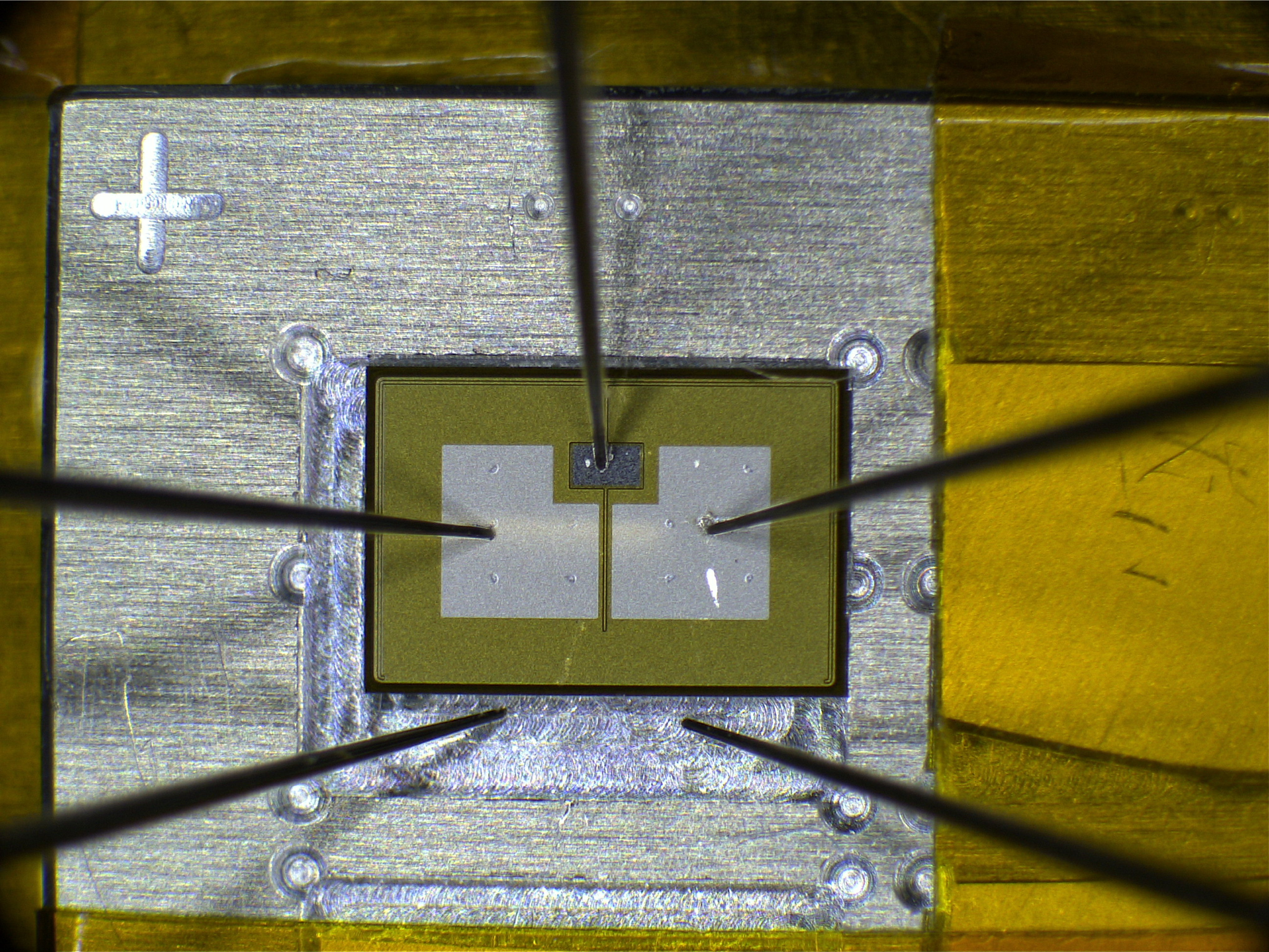}
    \caption{Chip under measurement on an aluminum chip holder.}
    \label{fig:chip-holder}
  \end{minipage}
  \hspace{4truemm}
  \begin{minipage}{0.67\linewidth}
    \centering
    \begin{minipage}{\linewidth}
      \centering
      \begin{minipage}[t]{0.1\linewidth}
        \centering
        \includegraphics[width=0.7\linewidth]{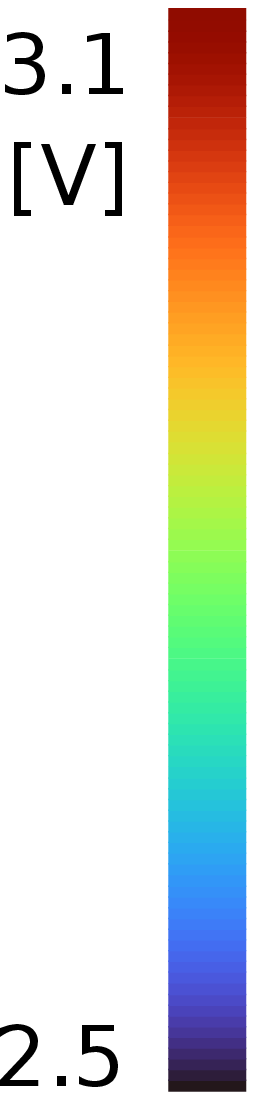}
      \end{minipage}
      \hspace{0pt}
      \begin{minipage}[t]{0.42\linewidth}
        \centering
        \includegraphics[width=\linewidth]{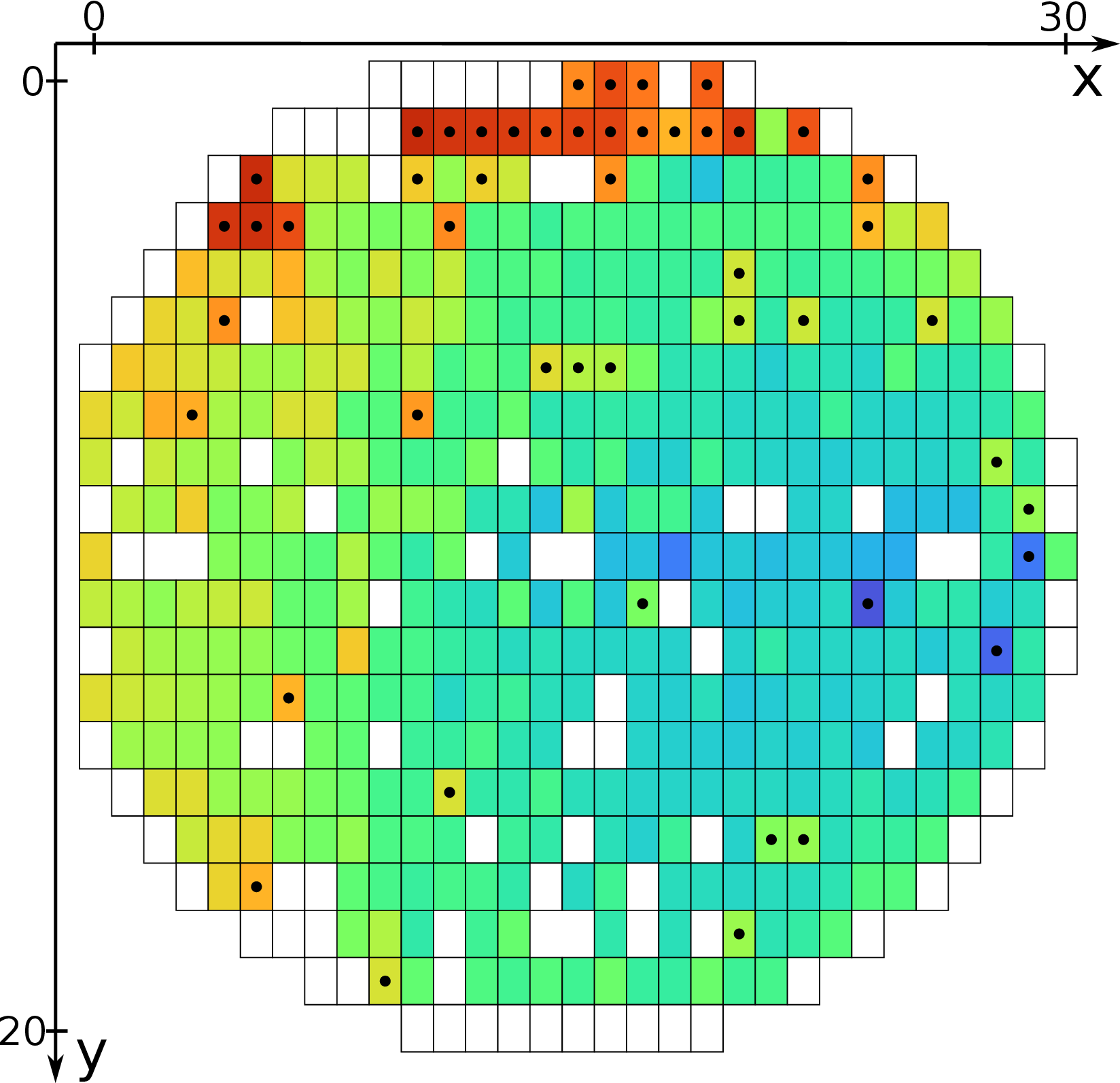}
       \caption*{(a) Measured $V_\mathrm{th}$ {\sato{distribution}}}
        \label{fig:vth-meas}
      \end{minipage}
      \hspace{0pt}
      \begin{minipage}[t]{0.42\linewidth}
        \centering
        \includegraphics[width=\linewidth]{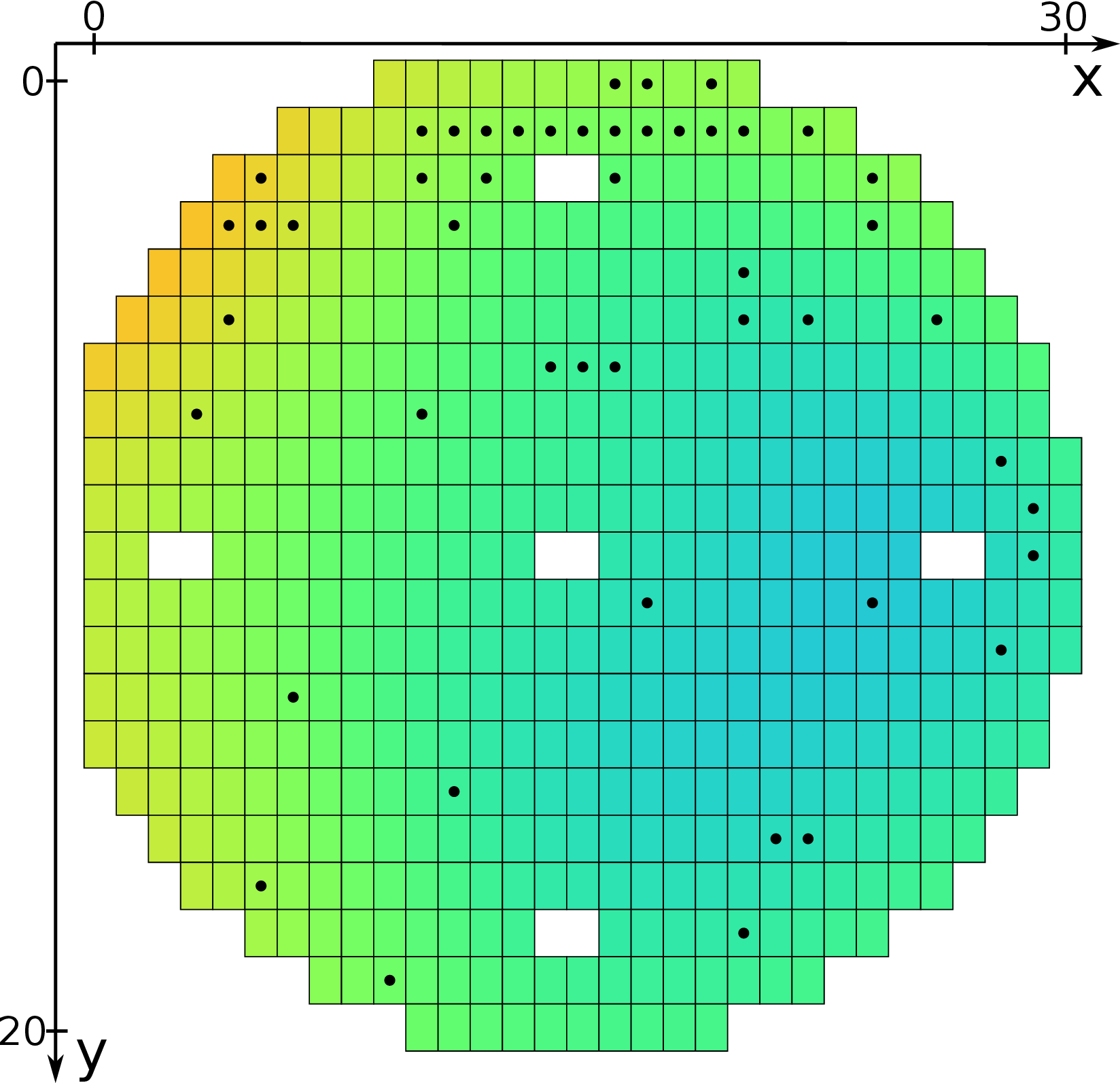}
        \caption*{(b) GPR mean of $V_\mathrm{th}$}
        \label{fig:vth-pred}
      \end{minipage}
    \end{minipage}
    \begin{minipage}{\linewidth}
      \centering
      \vspace{4truemm}
      \begin{minipage}[t]{0.1\linewidth}
        \centering
        \includegraphics[width=0.85\linewidth]{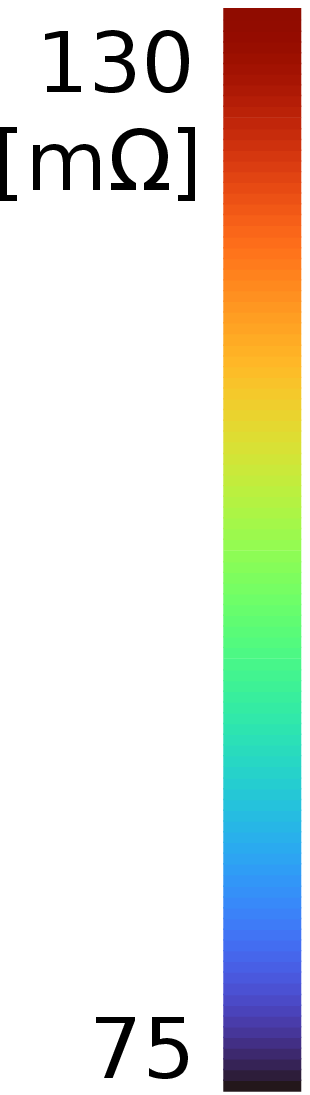}
        \vspace{0pt}
      \end{minipage}
      \hspace{0pt}
      \begin{minipage}[t]{0.42\linewidth}
        \centering
        \includegraphics[width=\linewidth]{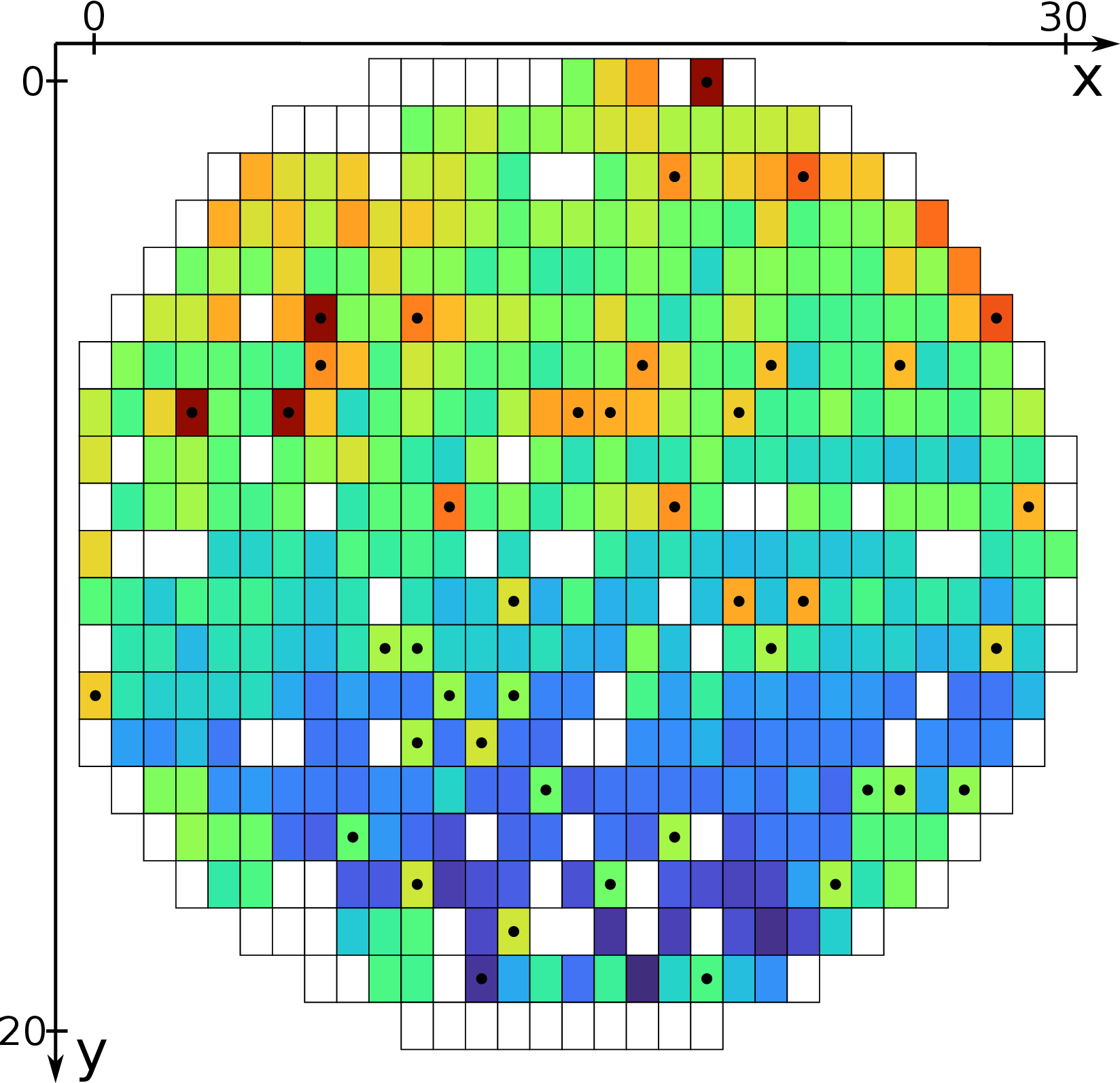}
        \caption*{(c) Measured $R_\mathrm{on}$ {\sato{distribution}}}
        \label{fig:ron-meas}
      \end{minipage}
      \hspace{0pt}
      \begin{minipage}[t]{0.42\linewidth}
        \centering
        \includegraphics[width=\linewidth]{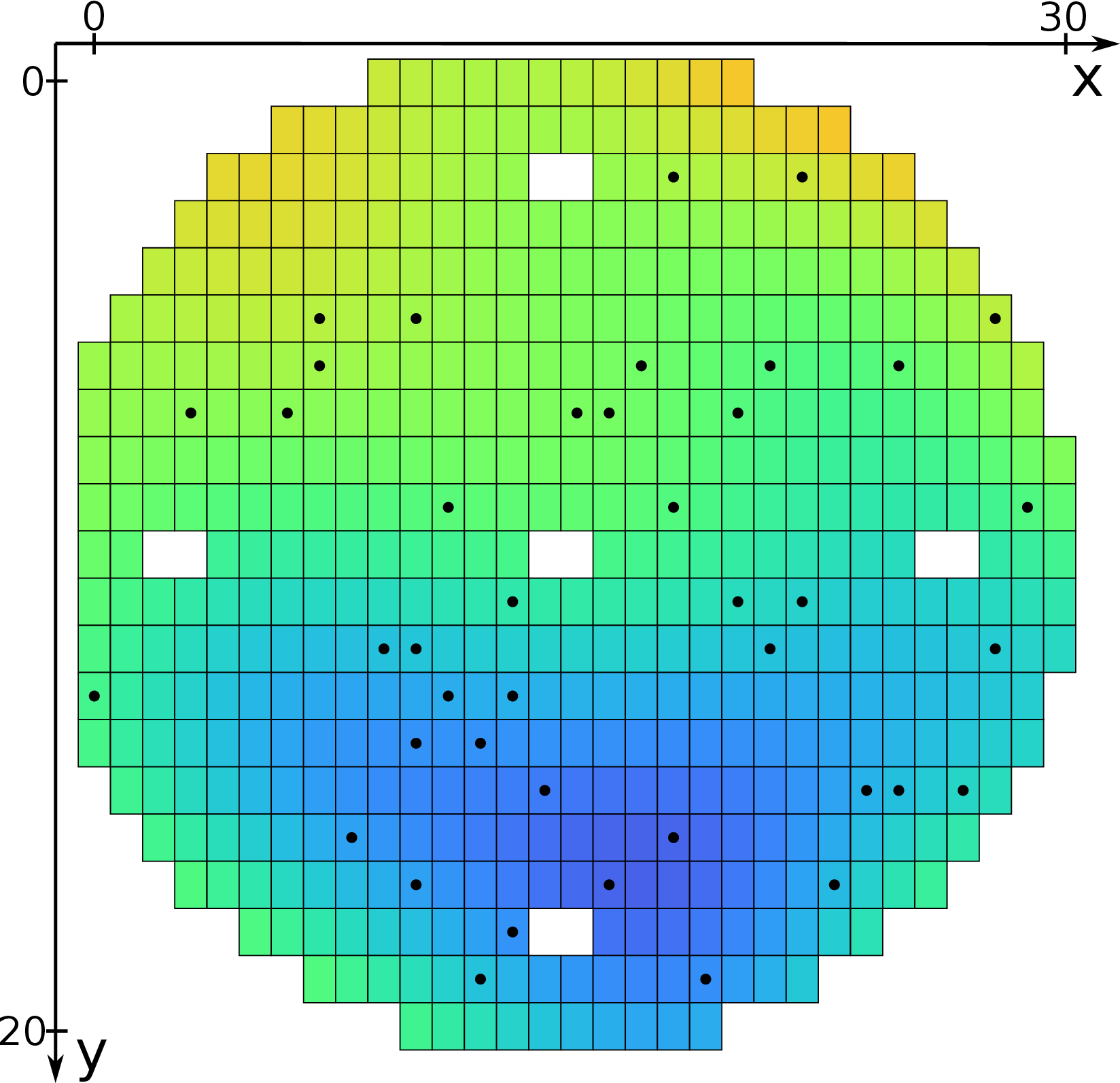}
        \caption*{(d) GPR mean of $R_\mathrm{on}$}
        \label{fig:ron-pred}
      \end{minipage}
      \caption{Wafer measurement results and the underlying trend {\sato{obtained}} as GPR mean.
      The chips with a dot are the outliers judged by the proposed method.}
      \label{fig:res}
    \end{minipage}
  \end{minipage}
\end{figure*}

\section{PROPOSED METHOD}

In this section, a method for detecting outliers using GPR with
Student-t noise is proposed.  As described {\sato{in the previous section}}, GPR is
a Bayesian non-parametric regression method that
yields a latent function with noise as a probability density.
The Student-t noise
is \shimo{included}
to alleviate the effects of outliers, i.e., to
avoid overfitting as compared to using standard Gaussian noise.  In the
proposed method, the spatial underlying trend of chip characteristics is
considered as a latent function $f$, {\sato{whose}} input $X$ is the
coordinate $(x, y)$ of a chip and {\sato{whose}} output with noise $Y$
is the characteristics of that chip.

The concept of the proposed method is {\sato{illustrated}} in
Fig.~\ref{fig:flow}.  The {\sato{measured value}} $\Y$ of chips in a
wafer follows an underlying spatial trend with some noise.  GPR is
applied to chip coordinate $\X$ and the measured values $\Y$, to
optimize the hyperparameters.  Then, the posterior distribution
$p(\Y^*)$ for each chip coordinate $\X^*$ is predicted by GPR.  Note
that the mean of the distribution $p(\Y^*)$ is just one of the
representations of the underlying trend, because the underlying trend is
a set of distributions $p(\f^*)$.  The probability of measured value
{\sato{to be}} in that distribution means the credibility of each chip
on the basis of the underlying characteristics trend.  From the set of
posterior distribution $p(\Y^*)$, the $100(1-\alpha)\%$ credible
interval can be calculated.  Here, $\alpha$ is called rejection rate.
\shimo{In the proposed method, the credible interval is used for the outlier detection.}
{\sato{By}} comparing the {\sato{measured values}} with their credible intervals,
each chip is classified as either outlier or not.
Fig.~\ref{fig:flow} {\sato{shows an example wherein $\alpha$ is set as 0.1}}.
Though the performances of chip A and chip B are close,
chip B should be considered as an outlier while chip A is not,
because {\sato{the performance of chip}} B excessively deviates from the underlying trend.
Chips like B are considered to potentially contain defects.
By adaptively determining the test limit for each chip on the basis of
the credible interval, the proposed method provides the optimal testing criteria
even when the large underlying trend is expected.

{\sato{More formally,}} the proposed {\sato{outlier detection}} method consists of the following five steps.
\begin{enumerate}
  \item Measure the performance of interest, such as on-resistance,
	breakdown voltage, threshold voltage, etc., of all available chips
	on a wafer.
  \item {\sato{Using}} the coordinates of chips $\X$ and {\sato{their}} measured characteristics $\Y$,
	the hyperparameters of GPR
	are optimized.
  \item For the $i$-th chip in the wafer, infer by GPR the probability distribution of $p(Y_i^*)$
    that $Y_i$ should follow, with the chip
    coordinate $X_i$ as $X_i^*$
  \item Judge the chip as an outlier if $Y_i$ is out of a
	$100(1-\alpha)\%$ credible interval. Otherwise, the chip is considered
	consistent with the underlying trend.
  \item {\sato{When there exist}} multiple performances of interest,
	{\sato{the rejection results are OR-ed.  In other words, a chip that is}} 
	judged to be an outlier {\sato{in}} at least one performance 
	should be {\sato{considered as an}} outlier.
\end{enumerate}

\section{MEASUREMENTS AND VALIDATION}
\label{sec:result}

In this section, the proposed method is validated and compared
{\sato{with}} the two conventional methods, DPAT and NNR.  {\sato{The
results}} for the measurement {\sato{data}} of a commercial SiC wafer
{\sato{are presented in Sec.~\ref{validation-meas}}}, and {\sato{the
results of}} {\sntn{an}} artificially generated demo dataset {\sato{are presented in
Sec.~\ref{validation-sim}}}.

\subsection{Measurement of commercial wafer}
\label{validation-meas}

The 435 chips on a commercial SiC wafer shown in Fig.~\ref{fig:s2301}
are measured.  {\sato{The photograph of a chip during probing is}} shown
in Fig.~\ref{fig:chip-holder}.
\shimo{In this study, the threshold voltage $V_\mr{th}$ and on-resistance $R_\mr{on}$ are
chosen as the performances of interest.}
{\sato{$V_\mr{th}$ is defined as the gate
voltage when}} $V_\mr{ds} = 20~\mr{V}, I_\mr{d} = 4.4~\mr{mA}$, and
$R_\mr{on}$ is measured at $V_\mr{gs} = 20~\mr{V}, V_\mr{ds} =
0.5~\mr{V}$.  {\sato{The temperature of the wafer is controlled}} at
40$\mr{^\circ C}$.


Figs.~\ref{fig:res}(a) and~\ref{fig:res}(c) show the measured
$V_\mathrm{th}$ and $R_\mathrm{on}$ {\sato{distributions}}.
The measured values of $V_\mr{th}$ range from 2.5 to 3.1~V,
and {\sato{those of}} $R_\mr{on}$ range from 75 to 130~$\mr{m\Omega}$.
Most of these values are still within the range of the official datasheet.
However, {\sato{chips having the}} values {\sato{significantly off}}
the underlying trend {\sato{present randomly}} on the wafer.
{\sato{In other words,}} 
those devices that exhibit significantly different characteristics
from their surrounding \shimo{trend} can be considered to
have long-term reliability issues.
The means of GPR posterior distribution are shown in
Fig.~\ref{fig:res}(b) for $V_\mathrm{th}$
and Fig.~\ref{fig:res}(d) for $R_\mathrm{on}$.
Here, the GPR \shimo{posterior} prediction is considered as the \shimo{underlying trend},
and regard the chips with their measured value exceeding 90\% credible interval
as outliers.
The detected outliers for each performance
are marked with a black dot in Fig.~\ref{fig:res}.
From the figure, it can be said that the proposed method successfully
detects outlier chips.  It is found that $V_\mr{th}$ follows the
estimated underlying trend well, {\sato{while}} $R_\mr{on}$ {\sato{suffers from}} a larger
random \shimo{component}.  As a result, there are as many outliers in
$V_\mr{th}$ as those in $R_\mr{on}$, because the prediction of
$V_\mr{th}$ has a narrower credible interval.

\begin{figure}[]
  \centering
  \begin{minipage}{\linewidth}
    \centering
    \begin{minipage}{0.48\linewidth}
      \centering
      \includegraphics[width=\linewidth]{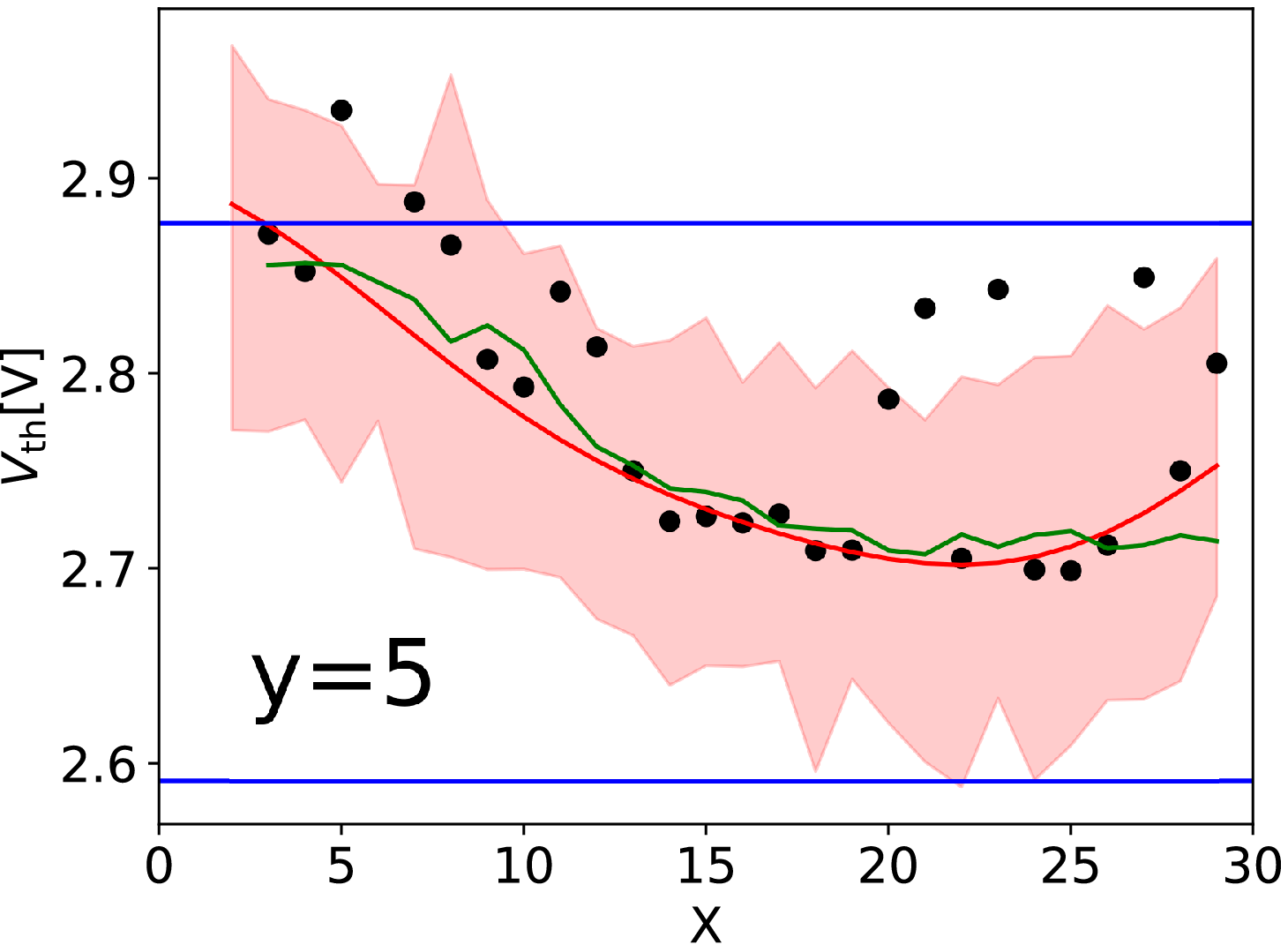}
    \end{minipage}
    \hspace{0.5truemm}
    \begin{minipage}{0.48\linewidth}
      \centering
      \includegraphics[width=\linewidth]{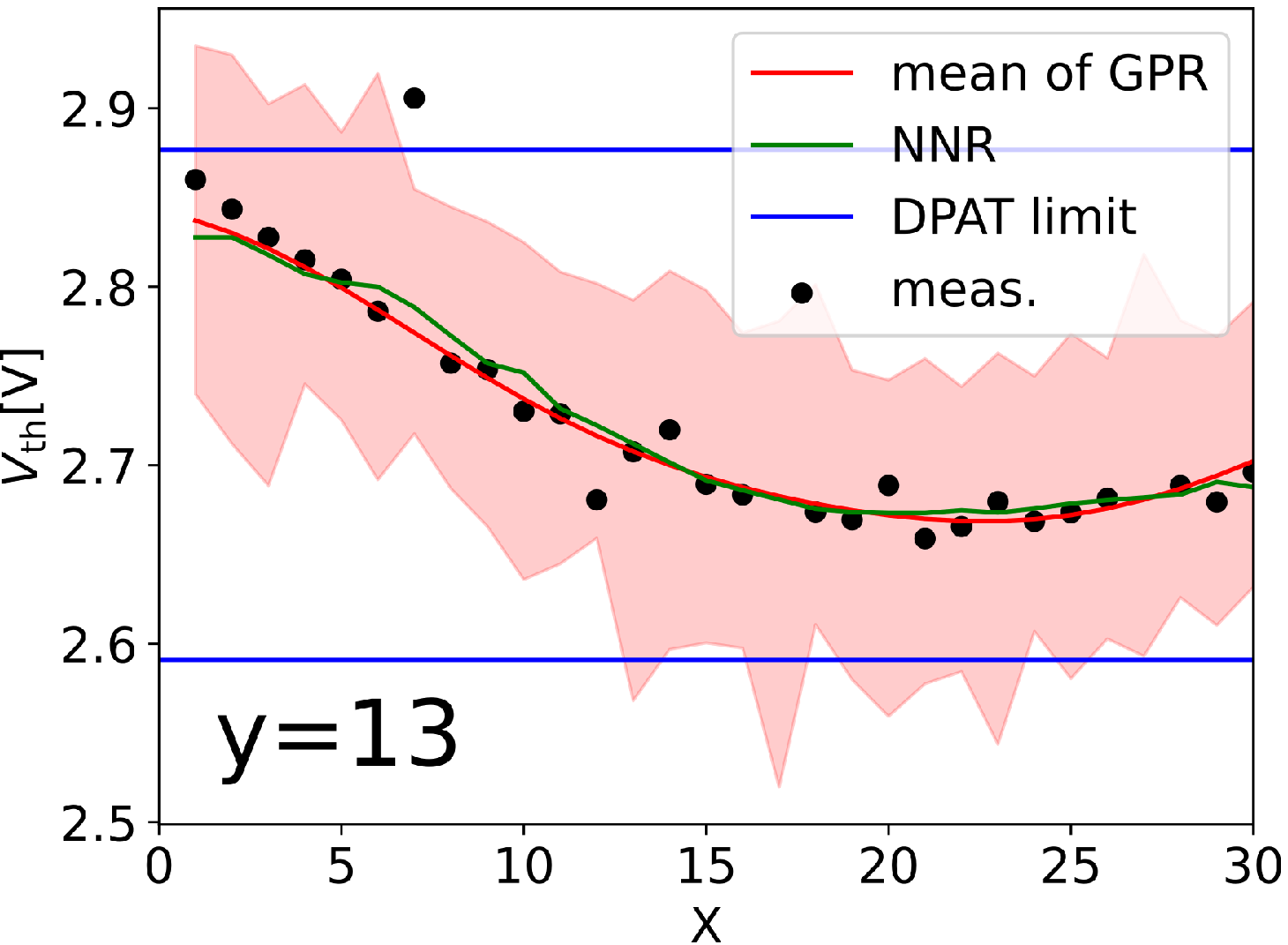}
    \end{minipage}
    \caption*{(a) $V_\mr{th} [\mr{V}]$}
  \end{minipage}
  \begin{minipage}{\linewidth}
    \vspace{4truemm}
    \centering
    \begin{minipage}{0.48\linewidth}
      \centering
      \includegraphics[width=\linewidth]{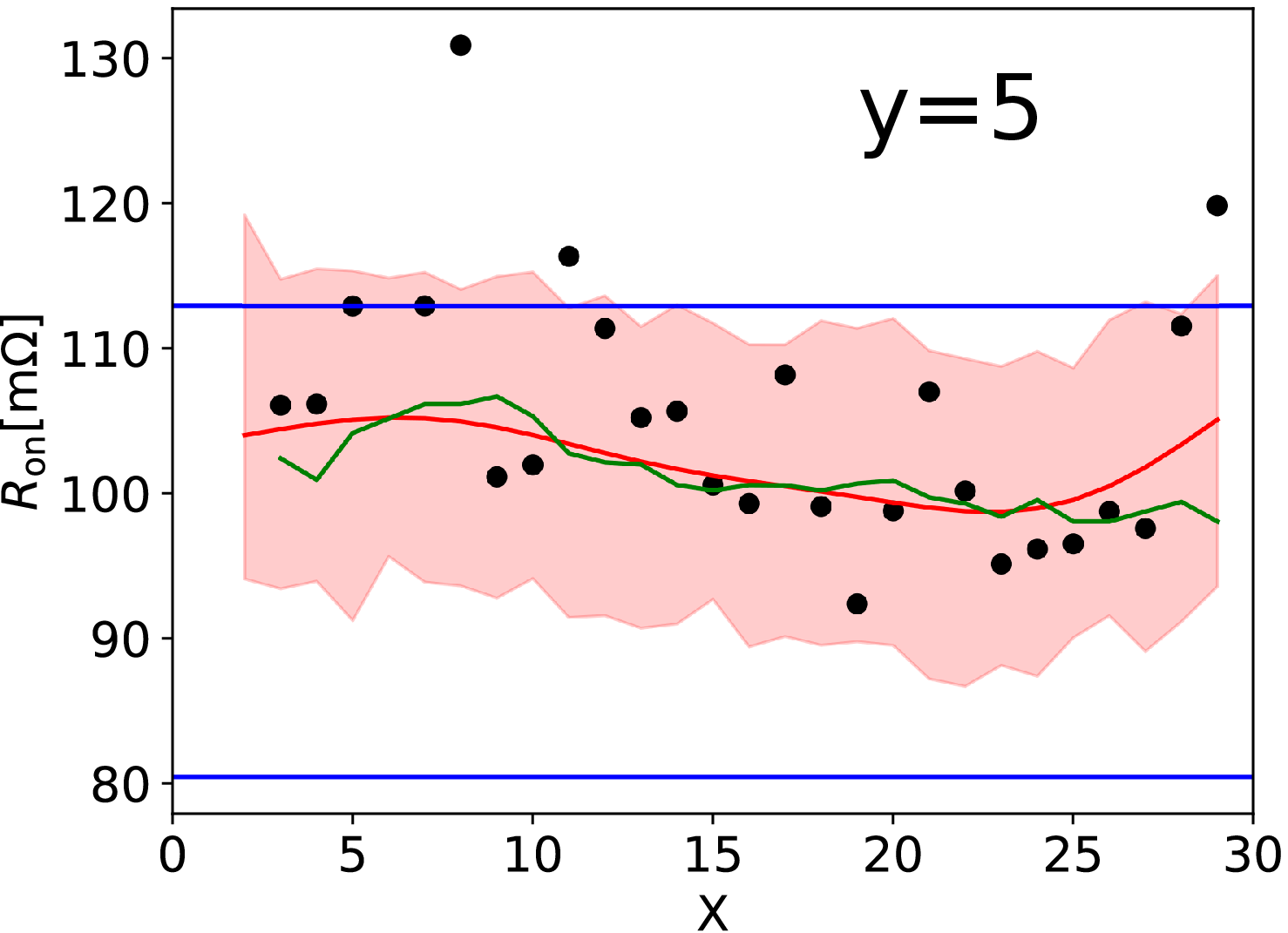}
    \end{minipage}
    \hspace{0.5truemm}
    \begin{minipage}{0.48\linewidth}
      \centering
      \includegraphics[width=\linewidth]{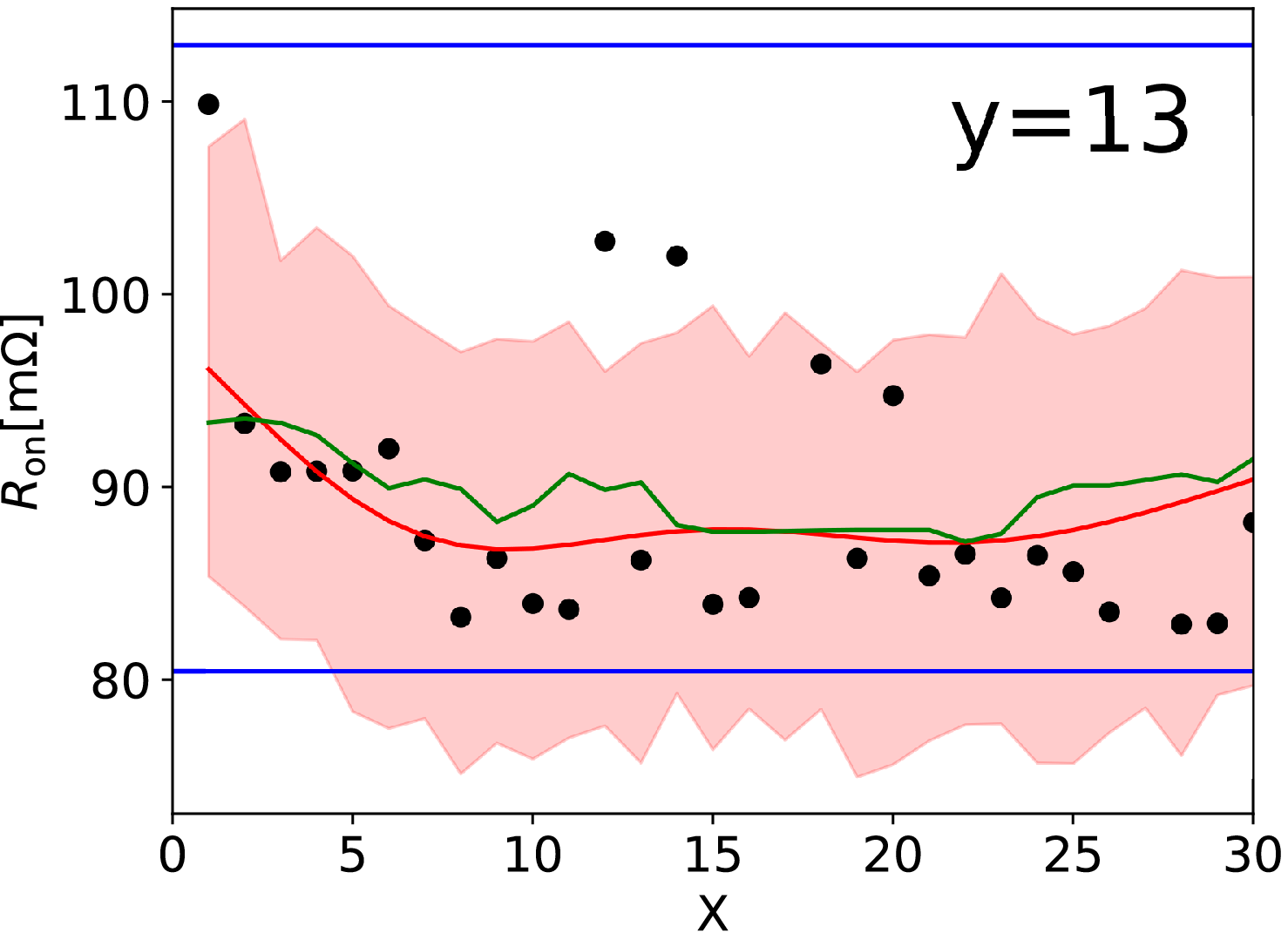}
    \end{minipage}
    \caption*{(b) $R_\mr{on} [\mr{m\Omega}]$}
  \end{minipage}
  \caption{Measurement versus the mean of GPR with 90\% credible interval at $y=5$ and $y=13$.
  The {\sato{trend given by NNR and the thresholds given by}} DPAT are also drawn.}
  \label{fig:dist}
\end{figure}

Fig.~\ref{fig:dist} {\sato{shows a comparison}} of the measurement and
the predicted distributions of GPR for the rows $y=5$ and 13.  In these
figures, in addition to the mean of the estimated distribution, the 90\%
credible {\sato{intervals are also depicted.}}  Regarding
$V_\mathrm{th}$ of $y=13$, most of the chips are located very close to
the estimated mean curve and have fallen within the credible interval,
indicating the \shimo{underlying trend} is dominant {\sato{over random
noise}}.  $R_\mr{on}$ of $y=5$ has a wider credible interval due to the
large random noise.  Meanwhile, though there are many outliers for
$V_\mr{th}$ of $y = 5$ and $R_\mr{on}$ of $y=13$,
the GPR posterior is not overfitted to them.

The 90\% (1.645$\sigma$) limit of DPAT and the underlying trend
predicted by NNR are also indicated in Fig.~\ref{fig:dist}.  Since DPAT
imposes the same limit for all the chips without taking into account the
underlying trends on the wafer, a large number of outlier chips
{\sato{with different characteristics from their neighbors}} are
classified as inliers.  This may lead to inappropriate testing results.
The NNR trend is not smooth compared to the GPR mean.
Since the residual from the NNR trend is often utilized to judge outliers,
the {\sato{unsmooth}} trend may cause wrong judgments.

\begin{figure}[]
  \centering
  \begin{minipage}{0.48\linewidth}
    \centering
    \includegraphics[width=\linewidth]{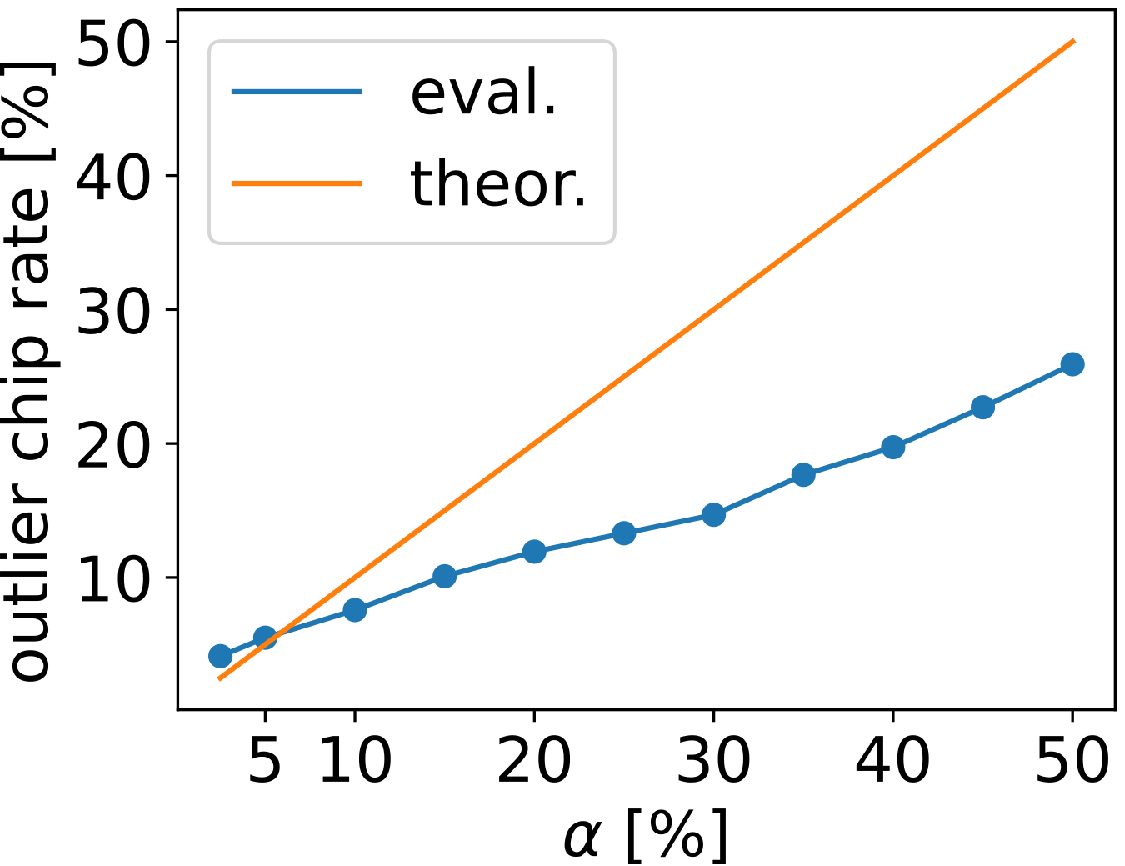}
    \caption*{(a) $V_\mr{th}$}
  \end{minipage}
  \hspace{1truemm}
  \begin{minipage}{0.48\linewidth}
    \centering
    \includegraphics[width=\linewidth]{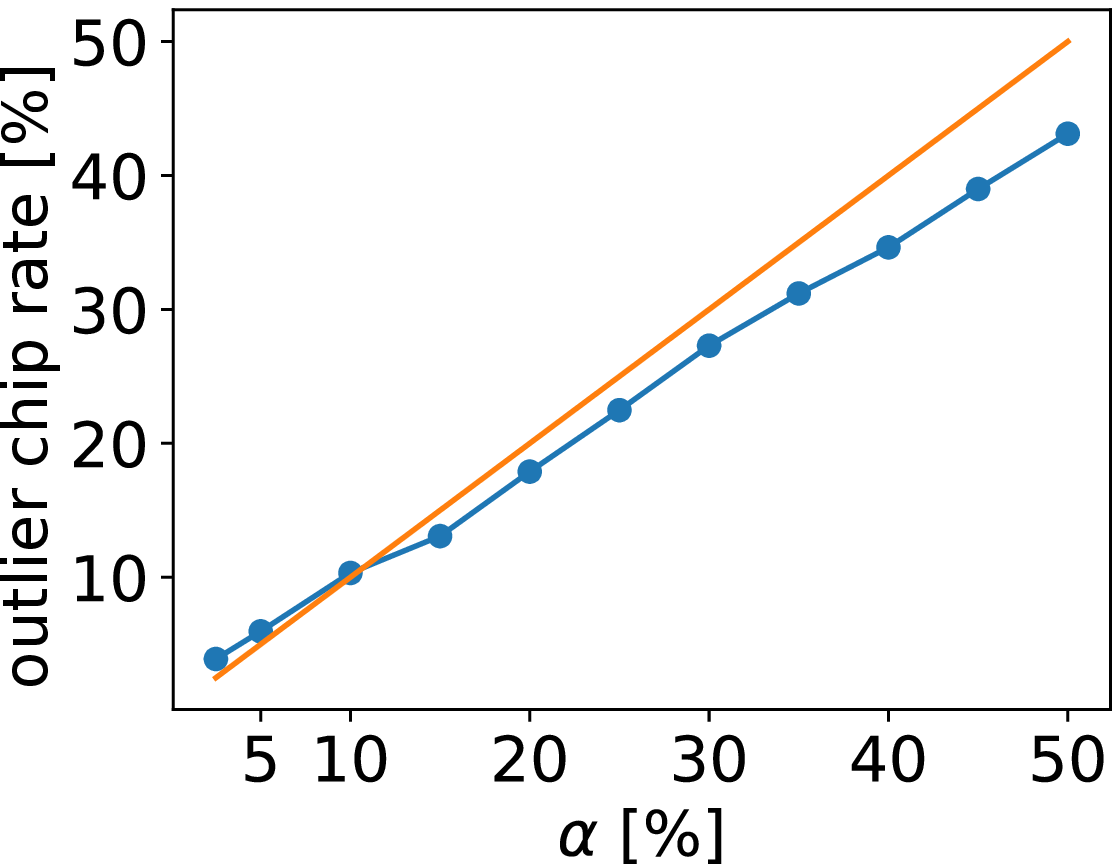}
    \caption*{(b) $R_\mr{on}$}
  \end{minipage}
  \caption{The relationship between $\alpha$ and {\shimo{outlier}} chip rate.
    The theoretical line where $\alpha$ is equal to the \shimo{outlier} chip rate is also drawn.}
  \label{fig:alpha}
\end{figure}

In the proposed method, the {\sato{fraction}} of {\sato{chips that are
judged as outliers is highly dependent}} on the rejection rate $\alpha$.
In order to reject potential outliers {\sato{close to the threshold}},
{\sato{a larger $\alpha$ may be used.}}
However, too large $\alpha$ {\sato{value will}} result in {\sato{a
lower}} yield, so it {\sato{must}} be set carefully.
Fig.~\ref{fig:alpha} shows the
\shimo{outlier chip rate {\sato{as a function of}} $\alpha$.
{\sato{Here, the outlier chip rate is defined as the ratio of}} chips that are
determined to be outliers out of all chips measured.}

{\sato{When}} $\alpha$ {\sato{is small, the outlier chip rate becomes higher
than what is expected by}} $\alpha$
because {\sato{there exist significantly bad performing chips that will
be}} judged outliers regardless of {\sato{the value of}} $\alpha$.
{\sato{In contrast, when}} $\alpha$ is {\sato{larger than 0.2}}, the
{\sato{outlier chip rate}} is less than {\sato{what is expected by}} $\alpha$.
\shimo{In that range of $\alpha$, less obvious outliers are found,
resulting in more yield loss.}
Therefore, the {\sato{intersection}} 
where $\alpha$ becomes equal to the outlier rate {\sato{can be used as a good
candidate to determine the appropriate}} $\alpha$ value.  {\sato{In this}} example,
$\alpha$ is defined {\sntn{at}} around {\sato{0.1}} for both
$V_\mathrm{th}$ and $R_\mathrm{on}$ characteristics.

\subsection{Virtual measurement dataset}
\label{validation-sim}

\begin{figure*}[]
  \centering
  \begin{minipage}{0.67\linewidth}
    \centering
    \begin{minipage}[t]{0.1\linewidth}
      \centering
      \includegraphics[width=0.7\linewidth]{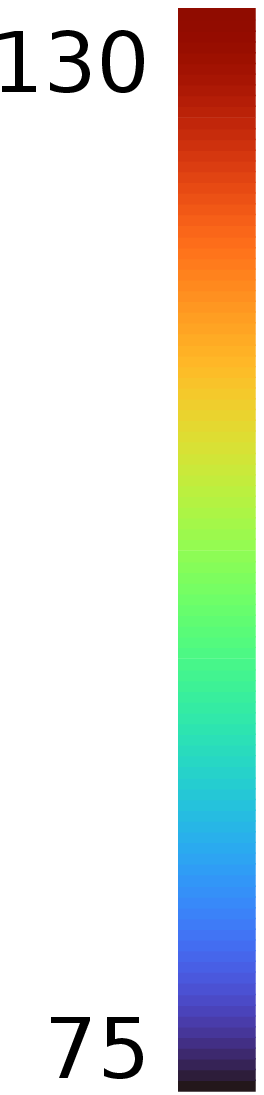}
    \end{minipage}
    \hspace{0pt}
    \begin{minipage}[t]{0.42\linewidth}
      \centering
      \includegraphics[width=\linewidth]{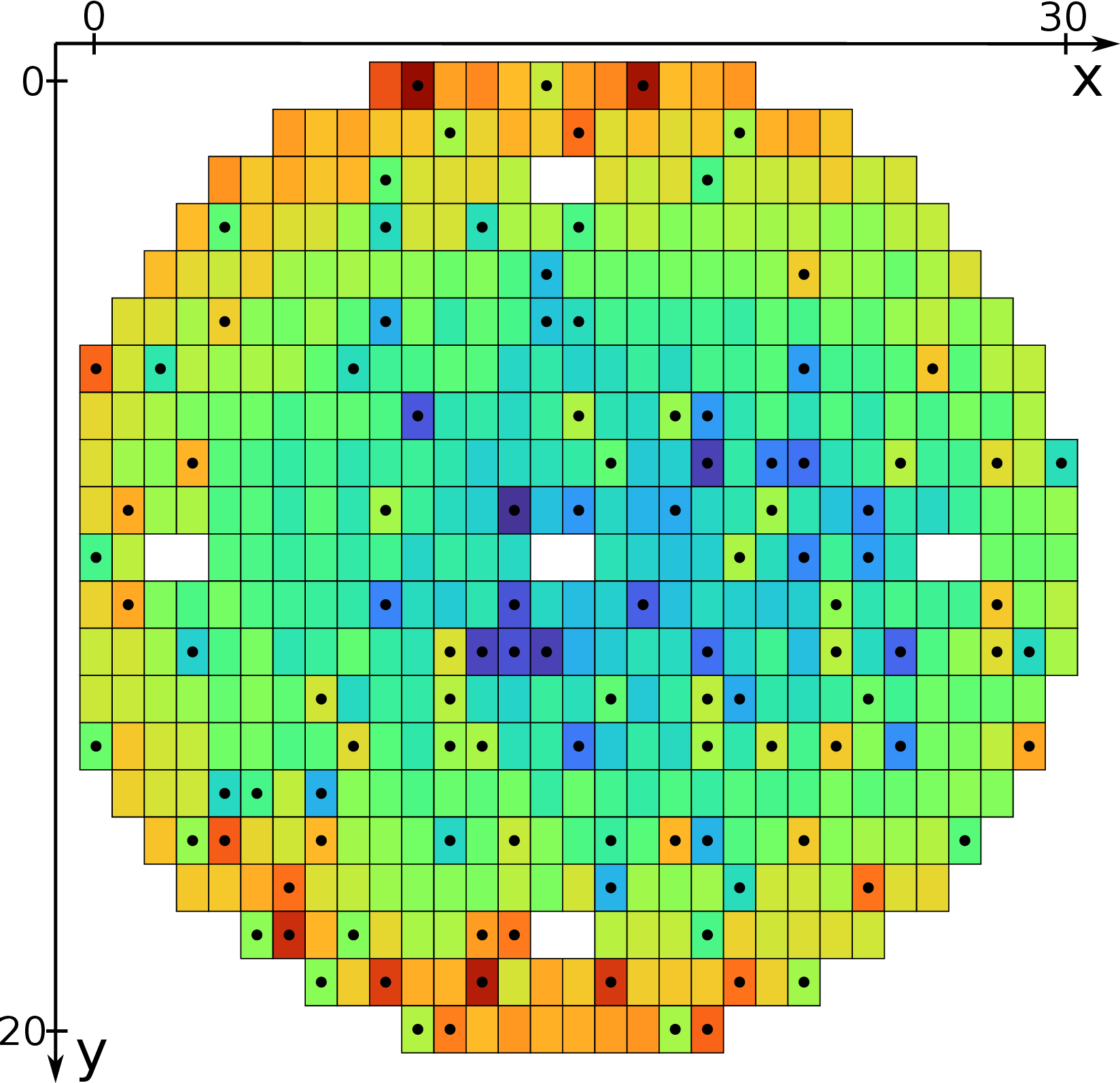}
     \caption*{(a) Performance {\sato{of virtual chips}}}
    \end{minipage}
    \hspace{0pt}
    \begin{minipage}[t]{0.42\linewidth}
      \centering
      \includegraphics[width=\linewidth]{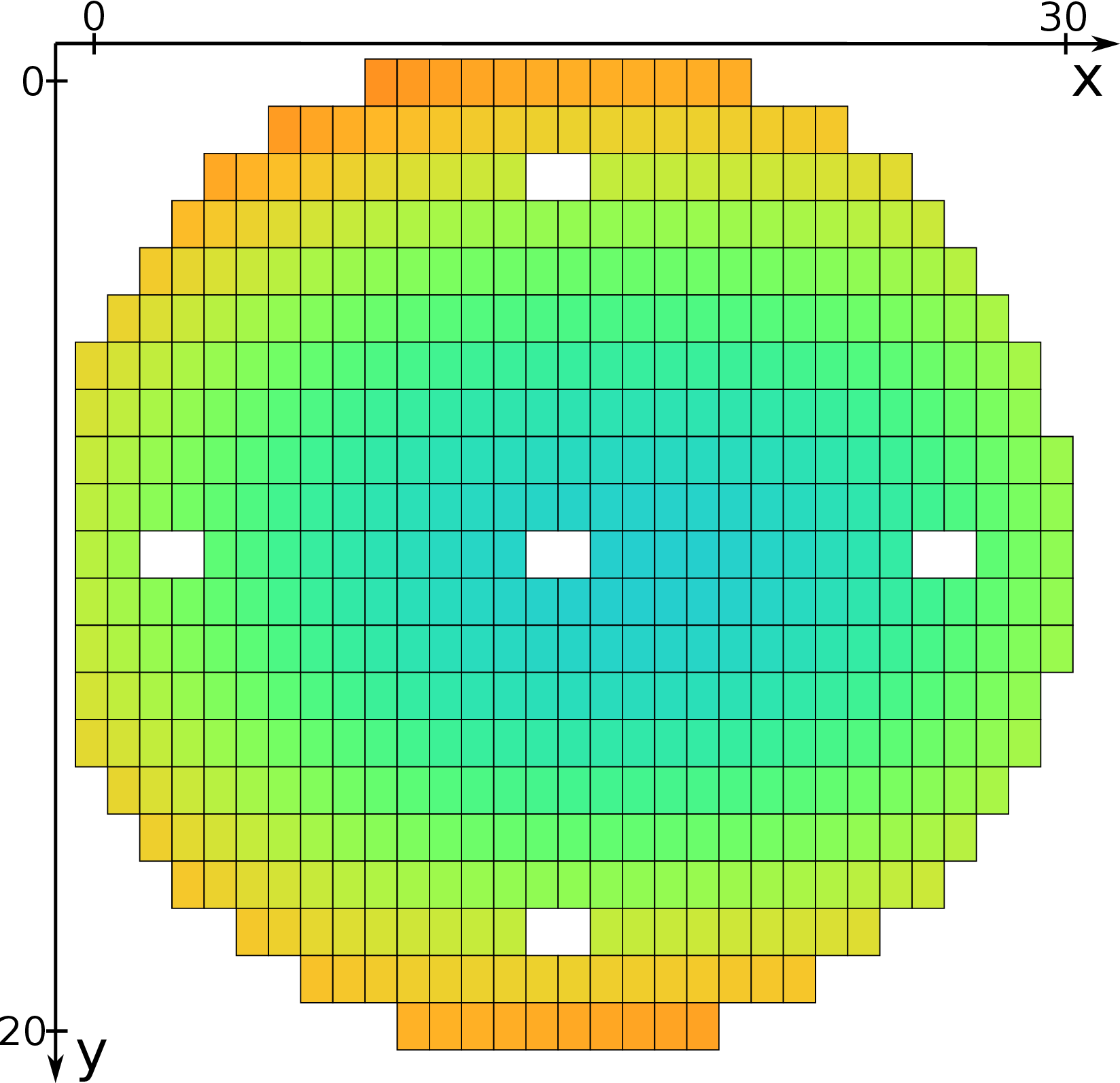}
      \caption*{(b) GPR mean}
    \end{minipage}
     \caption{{\sato{Visualization of a virtual measurement dataset}}.
      The chips with a dot are the outliers {\sato{determined}}
      by the proposed method.}
    \label{fig:demo}
  \end{minipage}
  \hspace{2truemm}
  \begin{minipage}{0.30\linewidth}
    \centering
    \includegraphics[width=0.9\linewidth]{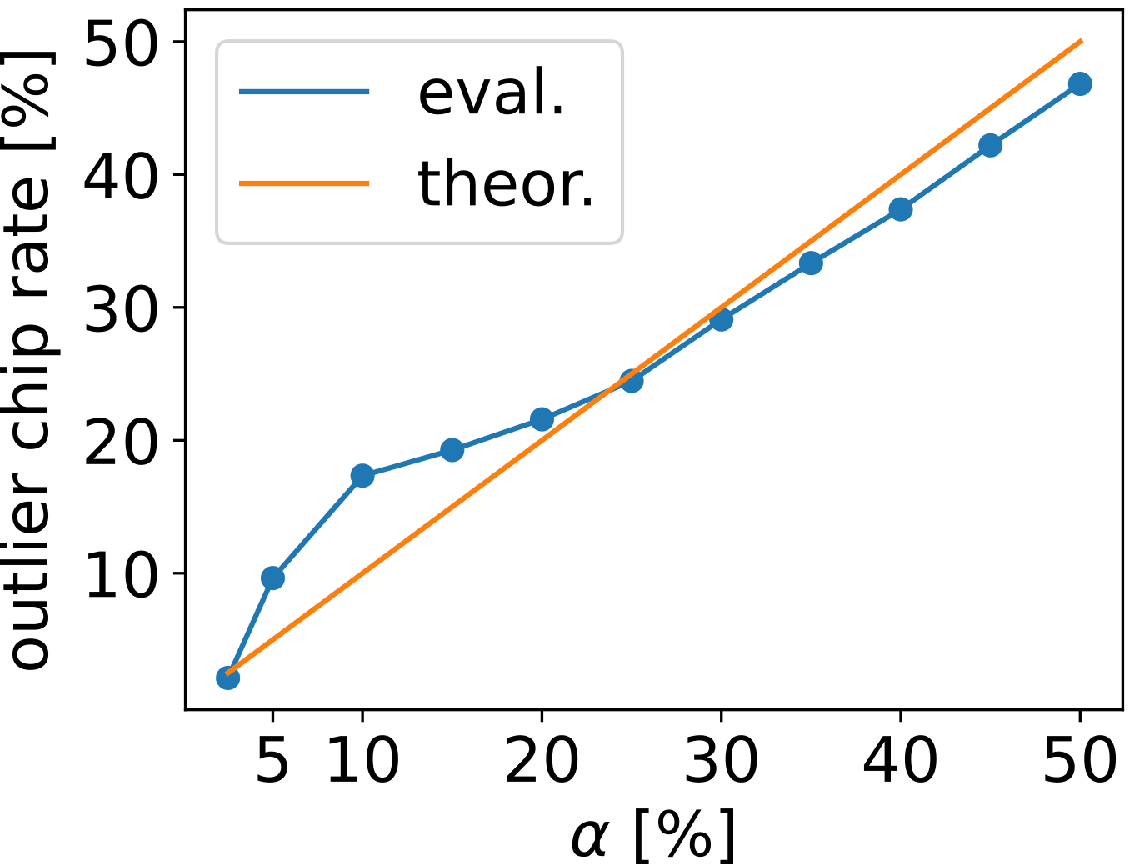}
    \caption{{\sato{The relationship between}} $\alpha$ and the
   {\shimo{outlier}} chip rate for the {\sato{virtual measurement}}
    dataset. {\sntn{The t}}heoretical line where $\alpha$ is equal to
   the {\shimo{outlier}} chip rate is also drawn.}
    \label{fig:alpha_demo}
  \end{minipage}
\end{figure*}

{\sato{In order to demonstrate the effectiveness of the proposed method,
a virtual}} dataset {\sato{consisting}} of 519 chips
on a wafer shown in Fig.~\ref{fig:demo}(a) is generated.
The {\sato{virtual}} performance data \shimo{$p(x, y)$} {\sato{has a baseline trend that}}
consists of {\sato{two components:}} global variation {\sato{component
$p_\mathrm{g}(x, y)$ as}} a function of the chip coordinate $(x, y)$
{\sato{on the wafer}} and random variation {\sato{component
$p_\mathrm{r}$ that is sampled from}} a normal distribution 
{\sato{that is mutually independent and identical for each chip}}.
{\sato{Here, the coordinates $x$ and $y$ are both integers.}} 
Additionally, {\sato{a relatively large random deviation \shimo{$p_\mathrm{d}$}
was taken into account}} for the 20\% of the entire chips.  {\sato{The
chips with this extra component added are defective and thus}} should be judged as {\sato{outliers}}.
{\sato{The overall performance of a chip is determined by:}} 
\begin{align}
 p(x, y) = p_\mathrm{g}(x, y) + p_\mathrm{r} + p_\mathrm{d},
 \label{eq:data}
\end{align}
where
\begin{align}
 p_\mathrm{g}(x, y) &= 90 + 0.06((x - 17)^2 + 4(y - 11)^2) \label{eq:glob} \\
 p_\mathrm{r} &\overset{iid}{\sim} \mca{N}(2.0, 1.0) \label{eq:rand} \\
 p_\mathrm{d} &\overset{iid}{\sim} \begin{cases}
    0 & (80\%) \\
    \mr{Uniform}(9, 11) & (10\%) \\
    \mr{Uniform}(-11, -9) & (10\%)
  \end{cases}. \label{eq:bad}
\end{align}

Fig.~\ref{fig:demo}(b) {\sato{shows}} the mean of the GPR posterior
distribution.  {\sato{The global trend given by Eq.~(\ref{eq:glob}) is
reproduced by the proposed method.  {\sntn{The t}}heoretical and GPR-based
{\shimo{outlier}} chip {\sntn{rates}} as functions of $\alpha$ are shown}} in
Fig.~\ref{fig:alpha_demo}.  From the {\sato{intersection}} in
Fig.~\ref{fig:alpha_demo}, the $\alpha$ is set to {\sato{0.2}}.
\begin{figure}[]
  \centering
  \begin{minipage}{\linewidth}
    \centering
    \begin{minipage}{0.48\linewidth}
      \centering
      \includegraphics[width=\linewidth]{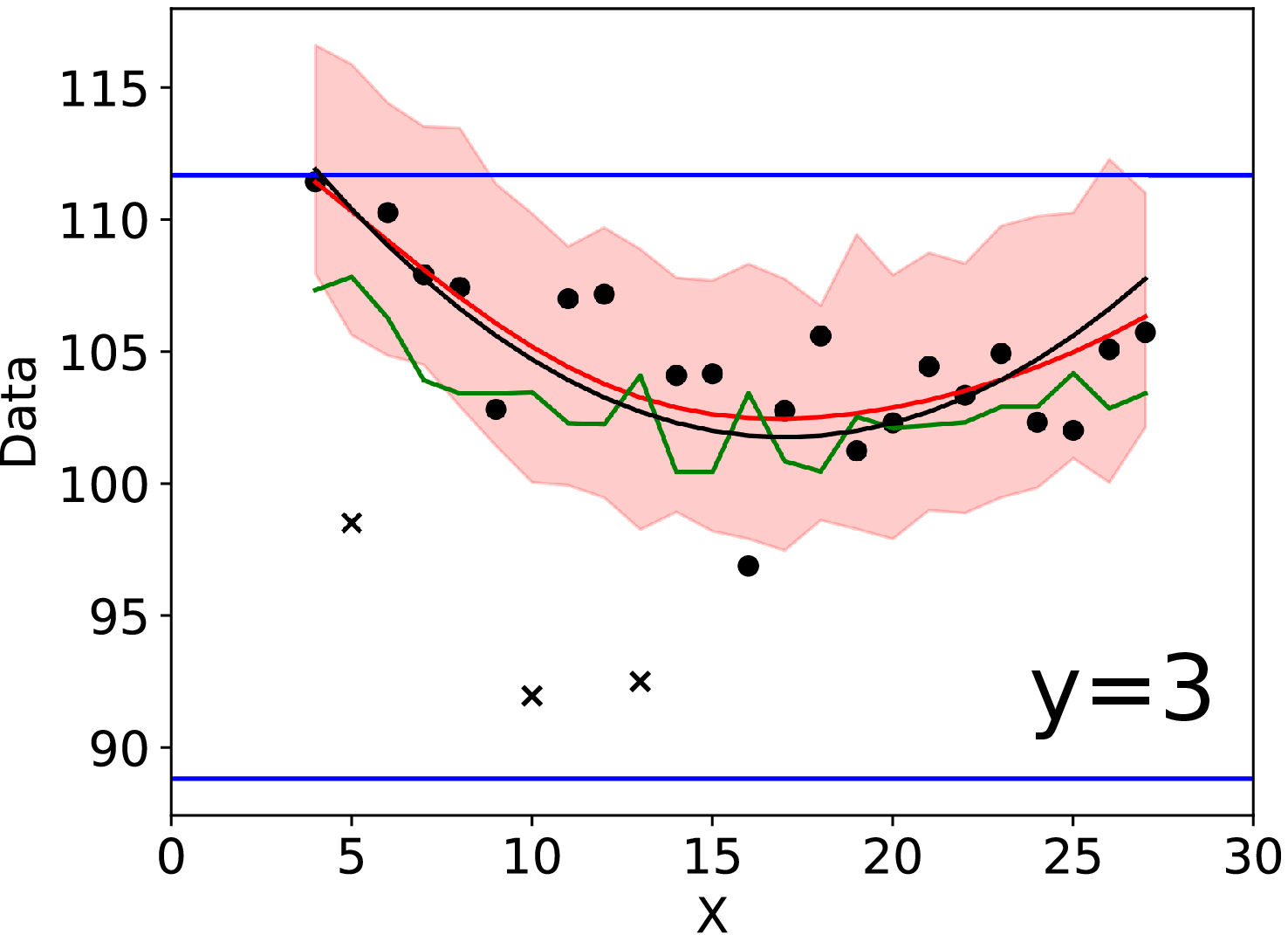}
    \end{minipage}
    \hspace{0.5truemm}
    \begin{minipage}{0.48\linewidth}
      \centering
      \includegraphics[width=\linewidth]{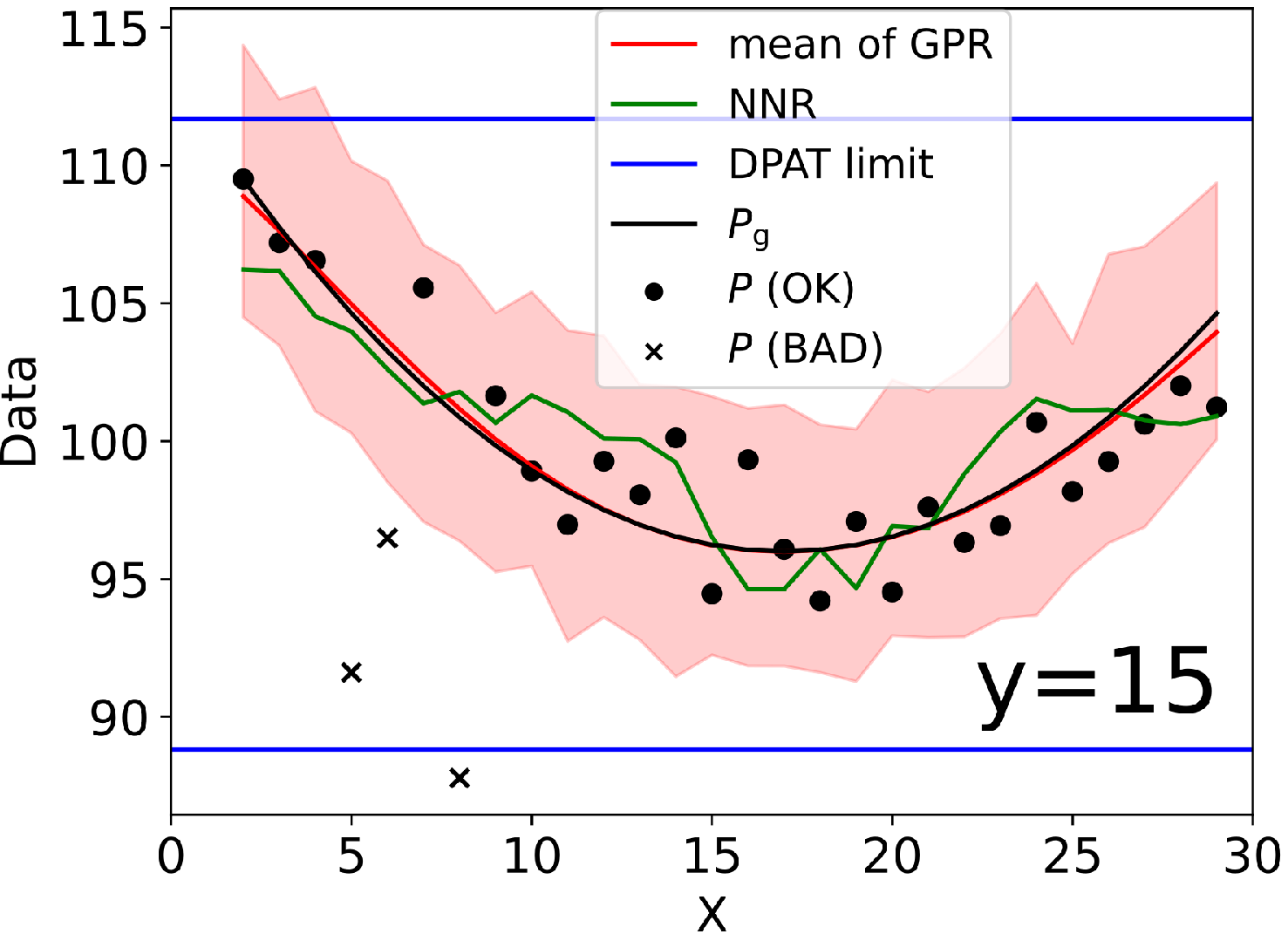}
    \end{minipage}
  \end{minipage}
  \caption{{\sato{Virtual measurement data versus the mean of}} GPR
  with 80\% credible interval at $y=3$ and $y=15$.
  The {\sato{trend given by NNR and the thresholds given by}} DPAT are also drawn.}
  \label{fig:dist_demo}
\end{figure}
{\sato{In}} Fig.~\ref{fig:dist_demo}, {\sato{the virtual data along
$y=3$ and $y=15$ is presented with the mean and the credible interval of
the GPR. Even with the presence of outliers,}} the proposed method
{\sato{extracts reasonably smooth global trend, \shimo{which is very close to
{\sato{the true global variation}} $p_\mr{g}$},
and provides a reasonable
threshold for separating good chips from defective ones.}}
The given {\sato{chips with different performance components are}} classified properly. {\sato{In contrast,}}
the {\sato{threshold given by the DPAT
is too broad and most of the
outlier chips have fallen in the ``pass'' region.  Similarly, the
background trend extracted by}} the NNR {\sato{method is different from
the ground truth due to the influence of outliers.}}
\begin{table*}[]
  \centering
  \begin{minipage}{0.8\linewidth}
  \centering
  \caption{Accuracy {\sato{comparison}} of outlier detection methods.
  Positive and negative {\sato{refer to}} outliers and inliers, respectively.
  Sensitivity is the {\sato{rate of outliers}} detected out of all given outliers.
  Specificity is the {\sato{rate of inliers}} detected out of all given inliers.
  {\sato{Higher}} score is better for both sensitivity and specificity.}
  \label{tab:accuracy}
  \begin{tabular}{|c||c|c|c|c||c|c|}\hline
    Method & True Positive & True Negative & \begin{tabular}{c}False Positive\\(yield loss)\end{tabular} &
      \begin{tabular}{c}False Negative\\(test escape)\end{tabular} & Sensitivity & Specificity \\\hline\hline
    This work & 102 & 407 & 10 & 0 & 1.00 & 0.98 \\\hline
    NNR & 89 & 395 & 22 & 13 & 0.87 & 0.95 \\\hline
    DPAT & 44 & 384 & 33 & 58 & 0.43 & 0.92 \\\hline
  \end{tabular}
  \end{minipage}
\end{table*}

{\sntn{The}} accuracy of each outlier detection method is {\sato{evaluated for
the virtual wafer in Fig.~\ref{fig:demo} and summarized}} in
Table~\ref{tab:accuracy}.  {\sato{The components of the confusion matrix
given by each method, sensitivity, and specificity are presented.
In terms of confusion matrix, the proposed method clearly outperforms
the two existing methods.  Accordingly,}} the sensitivity and
specificity {\sato{of the proposed method}} are better than that of NNR
and DPAT.  {\sato{In particular}}, the sensitivity {\sato{of the proposed
method}} is 1.00, meaning that all the given outliers are detected.  The
specificity {\sato{is}} 0.98, {\sato{which means that the rate of
misclassification}} of inliers as outliers is remarkably small.


\section{CONCLUSION}
In this study, an adaptive outlier detection
methodology based on the predicted posterior of GPR is proposed.
The proposed method successfully detected outliers
in {\sato{both}} experiments using {\sato{the measurement data of}} a commercial SiC wafer
and {\sato{virtual measurement}} dataset.
{\sato{In the experiments using the virtual measurement data,}} 
the accuracy of the proposed method {\sato{significantly outperformed the}} two conventional methods, NNR and DPAT.
Though $V_\mr{th}$ and $R_\mr{on}$ are selected {\sato{as}} the performance of interests,
the proposed method {\sato{can}} be {\sato{effective}} for {\sato{the tests using}} other characteristics,
such as breakdown voltage or parasitic capacitances.


\input{sub/ack.input_tex}

\bibliography{ref}

\end{document}